\documentclass[12pt]{article}
\pdfoutput=1
\usepackage[colorlinks=true,linkcolor=blue,urlcolor=blue,
  citecolor=blue]{hyperref}

\usepackage{fullpage}
\usepackage{amssymb,graphicx}
\usepackage{amsmath}
\usepackage[margin=0.5cm]{caption}
\usepackage{hyperref} 
\usepackage{cite}
\usepackage{upgreek}

\def\eps{\epsilon}
\def\d{\partial}
\def\l{\left(}
\def\r{\right)}

\newcommand{\be}{\begin{equation}}
\newcommand{\ee}{\end{equation}}
\newcommand{\bea}{\begin{eqnarray}}
\newcommand{\eea}{\end{eqnarray}}
\newcommand{\bg}{\begin{gather}}
\newcommand{\eg}{\end{gather}}
\newcommand{\bseq}{\begin{subequations}}
\newcommand{\eseq}{\end{subequations}}

\renewcommand{\ln}{\mathop{\rm ln}\nolimits}

\def\half{\frac{1}{2}}

\newcommand{\sij}{s_{ij}}
\newcommand{\thij}{\theta_{ij}}
\renewcommand{\Im}{\mathop{\rm Im}}

\def\mbf{\mathbf}
\def\A{\mathcal{A}}
\newcommand{\comment}[1]{}

\begin{document}
\baselineskip=15.5pt
\begin{titlepage}
\begin{center}
{\Large\bf Natural Tuning:\\
Towards A Proof of Concept\\ }
\vspace{0.5cm}
{ \large
Sergei Dubovsky, Victor Gorbenko, and Mehrdad Mirbabayi
}\\
\vspace{.45cm}
{\small  \textit{    Center for Cosmology and Particle Physics,\\ Department of Physics,
      New York University\\
      New York, NY, 10003, USA}}\\ 
\end{center}
\begin{center}
\begin{abstract}
The  cosmological constant problem and the  absence of new natural physics at the electroweak scale, if confirmed by the LHC,  may either indicate that the nature is fine-tuned or that a refined notion of naturalness is required. We construct a family of  toy UV complete  quantum theories providing a proof of concept for the second possibility. Low energy physics is described by a tuned effective field theory, which exhibits relevant interactions not protected by any symmetries and separated by an arbitrary large mass gap from the new ``gravitational" physics, represented by a set of irrelevant operators. Nevertheless, the only available  language to describe dynamics at all energy scales does not require any fine-tuning. The interesting novel feature of this construction is that UV physics is not described by a fixed point, but rather exhibits asymptotic fragility. Observation of additional unprotected scalars  at the LHC  would be a smoking gun for this scenario. Natural tuning also favors TeV scale unification.
\end{abstract}
\end{center}
\end{titlepage}

\section{Introduction and Summary}
For more than three decades the idea of naturalness \cite{tHooft:nat}
was the major guide for particle physics model building.
It is certainly premature to draw any definite conclusions from the LHC results at the moment. However, if there is one to be drawn, 
this would probably be that the physics at the electroweak scale appears tuned.
As a result of the 7(8)~TeV LHC run we have  solid experimental evidence for the existence of a scalar particle---the Higgs boson.  Shift symmetry for the Higgs is badly broken, with the Yukawa coupling to the top quark being the largest source of the breaking.  
Elementary scalars without shift symmetries are plagued with quadratic divergencies, which imply that in the presence of  {\it  generic} new physics, characterized by the energy scale $\Lambda_{NP}$, their mass acquires  corrections proportional to $\Lambda_{NP}^2$.
Gravity attests  the existence of new physics at least at the Planck scale. This tension is the essence of the electroweak hierarchy problem in the Standard Model. 

This argument provides an excellent motivation for a new physics at the TeV scale, which would cancel quadratic divergencies and screen the Higgs potential from large UV-dominated 
contributions. Observing such a physics at the LHC would be the most conservative resolution of the weak hierarchy problem. 
There is a number of well-motivated proposals for how it may happen. For instance,
quadratic divergencies would be eliminated if Higgs were a composite particle with the compositeness scale close to the electroweak scale \cite{Kaplan:1983fs,Kaplan:1983sm}. Alternatively,  cancelation of quadratic divergencies may come
from  loops of new perturbatively coupled particles.
Given that the largest quadratic divergence is associated with the top loop, it is natural to expect some of the particles responsible for cancellations to carry color, with stop being the classic example \cite{Dimopoulos:1981zb}. 

Both electroweak precision data and direct LHC constraints on new colored states
impose strong pressure on models of this kind. It still well may be that  the 14 TeV LHC run will discover natural TeV scale physics.
However, given the current stringent bounds, and waiting for the new LHC data to come, it is worthwhile to explore more adventurous possibilities. 
 
The second in the list of conservative options is  the possibility that the Higgs potential is tuned as a consequence of environmental anthropic selection in the vast landscape of (string) vacua with varying values of physical parameters, such as the Higgs mass and the vacuum energy \cite{Weinberg:1987dv,Agrawal:1997gf}. Similarly to the first scenario, this would not be something unseen before. It is absolutely uncontroversial that many of the physical parameters (such as the size of the Earth orbit) are selected environmentally.  Unfortunately, finding a direct evidence for the anthropic origin of the TeV scale is likely to be rather challenging although not necessarily impossible. (Mini)split supersymmetry is a good example of predictive model building inspired by these considerations \cite{ArkaniHamed:2004fb,Giudice:2004tc,Arvanitaki:2012ps,ArkaniHamed:2012gw}.

In the current paper we will explore a third possibility, which is more speculative than the previous two. If realized, unlike natural models and environmental selection, this will be 
a truly novel behavior, never observed in the past. This scenario is partially inspired by several earlier discussions in the literature, including \cite{Bardeen:1995kv,Shaposhnikov:2008xi,Farina:2013mla,Lykken} (see also \cite{Foot:2007iy,Heikinheimo:2013fta}).  At the most basic level it reduces to postulating that the fine-tuning problem is an artifact of the Wilsonian approach and Nature ``does not calculate" in the Wilsonian way, thus avoiding the fine-tuning associated with quadratic divergencies. We will refer to this proposal as ``natural tuning".
This cannot be a universally applicable approach and there are
many known instances, where Nature does follow Wilson in the choice how to calculate. To get convinced of the validity of this statement we may suggest the reader to look around and find a material  close to  a critical point associated with a second order phase transition. 

For the natural tuning to become more than a wishful thinking one should at least suggest an alternative language for calculating physical observables, which would not invoke fine-tuning. This will open the possibility for Nature to make an alternative choice.
In the current paper we make two concrete steps in this direction. First, we provide a specific formulation of the third scenario, directly in terms 
 of the properties of the renormalization group (RG) flow. Second, we construct a concrete two-dimensional realization of natural tuning, which may be considered as a proof of concept for this idea. 
 
 Interestingly,  our construction does more than providing an alternative language which does not exhibit fine-tuning in a situation which appears tuned in the Wilsonian description. At least at the current level of understanding, the non-Wilsonian language is the only one applicable at all energy scales  in the construction. Wilsonian description works at low energies and appears tuned.  At high energies it fails. If this feature survives further scrutiny, Nature does not have a choice in this class of models, if the goal is to cover all energy scales.
 
 To further specify  what we mean by natural tuning, it is instructive to formulate the hierarchy problem in more concrete terms.
 In particular, we prefer a formulation relying on invariant properties of the RG flow, without invoking the notion of quadratic divergencies whose presence is scheme dependent and may give the wrong impression that the hierarchy problem may be avoided by a careful choice of renormalization scheme, for instance by using the dimensional regularization. 

Our formulation  consists of two parts. The first one is completely conventional and its recent clear and refined presentation can be found, for instance, in \cite{Rattazzi:2008pe}. The second part defines the problem in a somewhat  more restricted way.
 Admittedly, it was designed as an {\it a posteriori} attempt to make sense of the ``solution" to be proposed later. However, we feel it matches well with what most people actually have in mind.
 
The necessary condition for the hierarchy problem to arise is the existence of at least two distinct physical scales. For concreteness, let us embed our discussion in the context of non-supersymmetric grand unified theories (GUT's), such as the $SU(5)$ model \cite{Georgi:1974sy}. Then the lower scale $\Lambda_H$ corresponds to the electroweak symmetry breaking, and the higher scale $\Lambda_{GUT}$ is the $SU(5)$-breaking scale. Almost by definition the physics between these two scales is approximately scale invariant and is governed by a nearby  conformal fixed point $CFT_{321}$, which in this case is simply the $SU(3)\times SU(2)\times U(1)$ Standard Model with all masses and couplings set to zero.    
This conformal theory is perturbed by a relevant operator ${\cal O}_H$, which is the Higgs mass, and at energies of order $\Lambda_H$ this perturbation goes strong. Below this scale the RG flow approaches another conformal theory $CFT_{31}$, characterized by the $SU(3)\times U(1)$ gauge group.
The $CFT_{321}$ theory is also perturbed by a set of irrelevant operators, originating from integrating out heavy GUT fields. From the viewpoint of the perturbed $CFT_{321}$ the scale $\Lambda_{GUT}$ sets the size of irrelevant perturbations.

In principle, already at this point one may wonder what protects the hierarchy $\Lambda_H\ll \Lambda_{GUT}$ given that no symmetry gets enhanced when the perturbation ${\cal O}_H$ is set to zero.
This is what we referred to as the conventional formulation of the hierarchy problem.
However, to visualize the origin of  the fine-tuning it is useful to make one additional step and to consider the situation as seen above the scale $\Lambda_{GUT}$. Dynamics there is well approximated by another conformal fixed point $CFT_5$, which is a free $SU(5)$ theory. This theory, in addition
 to ${\cal O}_H$, is perturbed by another set of  relevant operators   ${\cal O}^i_{GUT}$.  These correspond to masses of the $SU(5)$ adjoint Higgs and other possible GUT thresholds. All these deformations
  become non-perturbative at the scale $\Lambda_{GUT}$. Hence, from the viewpoint of the UV $CFT_5$ both $\Lambda_H$ and $\Lambda_{GUT}$
correspond to  energies where some relevant deformations go strong. 
\begin{figure}[t]
\begin{center}
\includegraphics[]{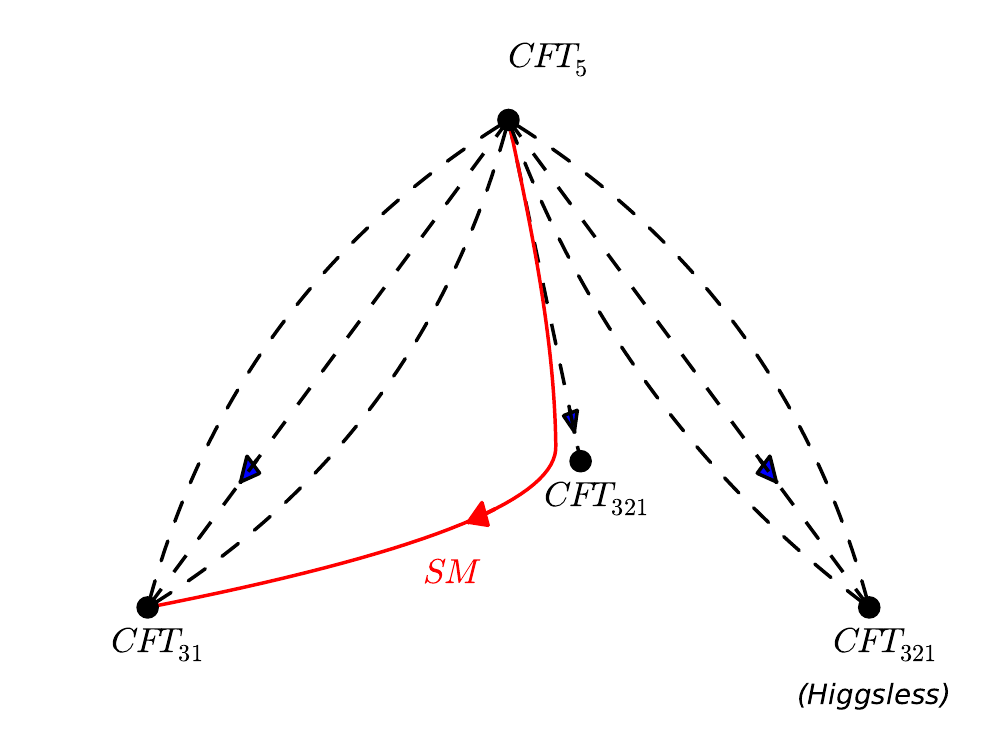}
\caption[]{\small{Fine-tuning in the Standard Model embedded in a non-supersymmetric $SU(5)$ GUT.} }

\label{fig:tuned}

\end{center}
\end{figure}

Now the fine-tuning is manifest (see Fig.~\ref{fig:tuned}). A generic direction in the space of possible relevant deformations of $CFT_5$ would correspond to a flow directly into $CFT_{31}$ or, for the opposite choice of the sign in front of ${\cal O}_H$, into a Higgsless version of $CFT_{321}$. The choice realized in nature corresponds to a very special trajectory which instead spends a huge amount of the RG time in the vicinity of $CFT_{321}$ and only then branches off into $CFT_{31}$, even though there is nothing special about this trajectory as seen in the vicinity of $CFT_5$ (which is equivalent to the statement, that there is no symmetry to protect the Higgs mass). 

It is apparent in this formulation that the problem persists  if a weakly coupled $CFT_5$ is replaced by a strongly interacting fixed point. 

However, the high energy scale associated with gravity does not fit in this picture. Indeed, 
a  UV complete quantum field theory is defined as an RG flow induced by deforming the UV CFT with a set of relevant operators, so that any scale may be described as the energy where some relevant (with respect to a UV CFT) deformation goes strong.
On the other hand, the Weinberg--Witten theorem  
\cite{Weinberg:1980kq} strongly suggests that a gravitational theory cannot be described by a conventional (even if strongly coupled) CFT in the UV, so that the Planck mass
$M_{Pl}$ may not have such an interpretation. 
Hence, one is tempted to speculate that the natural tuning may be realized in a situation when the high energy scale, such as $M_{Pl}$ for the Standard Model, is associated with gravity and cannot be characterized as a scale where some relevant operator becomes strong. In particular
 this implies that there should be no new massive thresholds as one approaches $M_{Pl}$. This makes it tempting to suggest that  only non-gravitational physics is required to be natural in the standard sense.

This scenario requires an asymptotic UV behavior of the RG flow, which is very different from what we usually encounter in field theory.
The whole discussion above would remain vacuous in the absence of a concrete construction illustrating this possibility. The main result of the current paper is to present such a construction.
The construction builds on and extends the results presented in \cite{Dubovsky:2012wk}.  Ref.~ \cite{Dubovsky:2012wk} presented a family of integrable two-dimensional models, exhibiting a surprisingly large number of properties expected from gravitational theories. A well known set of physical systems exhibiting this behavior is provided by  worldsheet theories of  infinitely long free critical strings (either bosonic or supersymmetric)\footnote{Note that the worldsheet theory of a free string is not a free theory.}.
Also they describe many aspects of  low energy dynamics of the QCD flux tubes  \cite{Dubovsky:2012sh,Dubovsky:2013gi} as observed on the lattice \cite{Athenodorou:2010cs}.

For the purpose of the current discussion, the crucial property of these models is that in spite of being UV complete theories, {\it i.e.} possessing a well-defined $S$-matrix describing
scattering at all energies, they do not correspond to  RG flows between  UV and IR CFT's. 
At low energies they appear indistinguishable from conventional quantum field theories and can be described by an IR CFT perturbed by a set of irrelevant operators. At high energies perturbation theory breaks down, however, there is no associated UV fixed point. The corresponding UV behavior was called asymptotic fragility. Currently, we do not have a good description of this behavior in the Wilsonian RG language. One of its characteristic features is the absence of well-defined local observables, which is expected to be a generic property of gravitational theories. This makes it hard to perturb
the corresponding flows. Any local perturbation breaks the finiteness of the scattering amplitudes, which explains the name.

The main technical result of the current paper is the extension of this construction to non-integrable quantum field theories in two dimensions. We will refer to this procedure as ``gravitational dressing" and it works in the following way. One starts with an arbitrary UV complete quantum field theory, characterized by a Lagrangian ${\cal L}(\psi)$, where $\psi$ is a set of matter fields. 
To avoid conventional field theoretical
hierarchy problem we take this theory to be natural.
However, in accord with the above discussion, this does not imply the absence of quadratic divergencies. A natural quantum field theory is allowed to exhibit unprotected scalars, as long as their mass sets the highest energy scale\footnote{More precisely, in principle there could be higher scales, but they should reside in other sectors of the theory, which are coupled sufficiently weakly to  the unprotected scalars.}. 

Let us denote by $S_n(p_i)$ the full set of the corresponding scattering amplitudes. Here $n$ stands for the total number of particles in the process, and $p_i$ is the set of real on-shell momenta, all of which are taken as incoming.
 In principle, these amplitudes contain all the information about the theory. Amplitudes $\hat{S}_n(p_i)$ of the gravitationally dressed theory are defined as
  \be
\label{Sdressing}
\hat{S}_n(p_i)=e^{i\ell^2 \!/4\sum_{i<j}p_i*p_j}S_n(p_i)\;,
 \ee
where 
\be
\label{skew}
p_i*p_j=\epsilon_{\alpha\beta}p_i^\alpha p_j^\beta\;,
\ee
and we made use of the fact that there is a natural cyclic order (according to their rapidities) on the space of non-zero momenta in two dimensions. 
The length scale $\ell$ is the analogue of the Planck length.
As we argue in detail later, gravitationally dressed amplitudes $\hat{S}_n(p_i)$ satisfy all the properties expected from a well-to-do $S$-matrix. The expression (\ref{Sdressing}) is reminiscent of amplitudes in non-commutative field theories. However, there are two important differences in our case. First, gravitational dressing is defined directly for the full $S$-matrix, rather than at the tree level, as it is usually done in non-commutative theories. Second,
we do not sum together exponential phase factors associated with all possible permutations of the particle momenta, but instead make use of the natural cyclic order on the set of momenta. This is fully consistent with the crossing symmetry in two dimensions and allows us to avoid  
acausalities associated with conventional space-time non-commutativity \cite{Seiberg:2000gc}.

At low energies a gravitationally dressed theory admits Wilsonian description in terms of the original Lagrangian deformed by an infinite number of operators suppressed by the positive powers of $\ell$. At high energies derivative expansions breaks down and the theory enters in the regime of asymptotic fragility.
There are several ways to see that it cannot be described by a conventional  CFT \cite{Dubovsky:2012wk}. In particular, holomorphic factorization on an infinite plane implies that the scattering amplitudes between left- and right-movers become trivial as one approaches the conformal limit, which is not the case in (\ref{Sdressing}).
 
This procedure provides a proof of concept for natural tuning. If the original quantum field theory possesses unprotected scalars,  the corresponding gravitationally dressed theory appears tuned at low energies, when it admits the Wilsonian description, provided the UV scale $\ell^{-1}$ is chosen to be much higher than the scalar masses.
However, gravitational dressing does not change the spectrum and we did not need to tune anything to construct $\hat{S}$.  
 
 One may argue that we are cheating here, because one can never see tuning directly at the $S$-matrix level. However, in general we are not aware of the procedure to construct $S$-matrices directly without going  through the Lagrangian\footnote{String theory provides a notable exception. See a further discussion  in the concluding Section~\ref{conclusions}.}, and that is where the tuning enters. If one succeeded in developing such a procedure, in our opinion, this should already be considered as a legitimate concern about the conventional idea of naturalness. In our view the example provided here goes further. 
 
 First, asymptotic fragility {\it forces} one to switch to the $S$-matrix  description, and this is expected to be a generic property of gravitational theories, which lack well-defined local observables. 
 
Second, in our discussion above we  introduced two notions of naturalness. The broader and the more conventional one says that \\
(I)  {\it If a natural theory
possesses unprotected relevant operators ({\it i.e.}, scalar masses), 
the corresponding energy scale should be the highest in the theory.}\\
The second, more restricted formulation, requires instead that\\
(II) {\it Among all possible scales set by relevant operators unprotected operators should correspond to the highest scale.}

Both formulations allow for an obvious generalization, accounting for the possibility of several weakly coupled sectors. Also it is straightforward to refine both formulations to account properly for the relevant operators which are close to  marginality along the lines of \cite{Strassler:2003ht,Luty:2004ye,Rattazzi:2008pe}.

 In conventional quantum field theories the two definitions agree, because one can always classify operators from the point of view of the ultimate UV CFT, as illustrated in Fig.~\ref{fig:tuned}, so that {\it every} scale can be interpreted as a scale corresponding to a certain relevant deformation of the UV fixed point.
On the other hand, as the example of gravitational dressing illustrates,  this is not true in general. The scale $\ell^{-1}$ does not have such an interpretation. It is tempting to speculate 
that for gravitational theories this situation may be generic and the second formulation is preferred.

 Finally, another interesting proposal, named ``classicalization", relating the electroweak hierarchy problem to the possibility of a non-Wilsonian UV completion has been put forward recently \cite{Dvali:2010jz}. One may ask whether natural tuning is related to classicalization.
 The answer is that the two ideas are very different both in their approach to the hierarchy problem and in the scenario for the UV completion. Unlike natural tuning, all  scenarios relating classicalization to the origin of the weak scale, as discussed in \cite{Dvali:2010jz},
 imply new natural physics close to the TeV scale. As far as the dynamics of UV completion goes, as proven by the authors of classicalization \cite{Dvali:2012zc}, the classical Nambu--Goto action with a single transverse field (which was referred to as the Dirac--Born--Infeld action in \cite{Dvali:2012zc}) with  a ``subluminal" sign (in the sense of \cite{Adams:2006sv}) does not allow classicalization. Instead, it provides the first and the simplest example of an asymptotically fragile theory.

The rest of the paper is organized as follows. In section~\ref{shocks} we briefly introduce integrable asymptotically fragile theories.
To minimize the overlap with \cite{Dubovsky:2012wk} we organize our discussion from one particular viewpoint, which was mentioned, but not emphasized in  \cite{Dubovsky:2012wk}.
Namely, these models can be thought of as describing scattering of gravitational shock waves. This naturally leads us to the holographic representation of the corresponding scattering amplitudes inspired by the derivation of the  gravitational eikonal scattering presented in \cite{Verlinde:1991iu}. In section~\ref{dressing} we demonstrate that this holographic description allows for a natural extension, covering also non-integrable case and resulting in the gravitational dressing, as presented in (\ref{Sdressing}). We discuss unitarity, crossing symmetry, and analyticity of the dressed $S$-matrix, and present several tree level and one-loop calculations illustrating how it can be reconstructed perturbatively, order-by-order in the small $\ell$ expansion, from a local effective field theory action. In the concluding section~\ref{conclusions} we briefly discuss some implications for the LHC model building, if one takes natural tuning seriously.

\section{Gravitational Shock Waves and Strings}
\label{shocks}
Let us consider gravitational scattering of relativistic point-like particles in the trans-Planckian regime, $E\gg M_{Pl}$, and large impact parameter $b\gg R_s$, where
\[
R_s=\l{E\over M_{Pl}^{d-2}}\r^{1\over d-3}
\]
is the Schwarzschild radius corresponding to the center-of-momentum energy $E$, and $d\geq 4$ is the space-time dimensionality. It is well-known that in this regime scattering can be described using classical Einstein gravity. At large impact parameter the process is elastic and the deflection angle is small. 
The dominant effect of scattering is the eikonal phase shift, which can be calculated either by resumming cross-ladder diagrams \cite{Amati:1987wq,Amati:1987uf}, or by solving for classical geodesics in a shock wave geometry \cite{Aichelburg:1970dh} produced by one of the particles \cite{'tHooft:1987rb}. Yet another approach, which turns out to be especially instructive in what follows, reduces the problem to a straightforward topological field theory calculation \cite{Verlinde:1991iu}. All these techniques result in the following answer for the forward scattering phase shift,
\be
\label{eikonal}
e^{i2\delta_{eik}(s)}=e^{i\ell^2 s/4}\;.
\ee
Here $s$ is the Mandelstam invariant and for gravitational scattering at $d> 4$ one finds
\[
\ell^2\propto G_Nb^{4-d}\;,
\]
where $G_N$ is the Newton constant. At $d=4$  one needs to introduce the IR cutoff $R_{IR}$, so that $\ell^2\propto G_N\log R_{IR}/b$. 

As observed in \cite{Dubovsky:2012sh},\cite{Dubovsky:2012wk}, the phase shift of the form (\ref{eikonal})  can be used on its own to define an integrable reflectionless massless theory in two dimensions.
Any integrable two-dimensional theory is completely determined by its two-to-two scattering $S$-matrix, which in this case takes the form
\be
\label{GGRT}
S_{GGRT}=e^{i2\delta_{eik}}{\bf 1}\;.
\ee
Here ${\bf 1}$ is a unit matrix in a two-particle flavor and momentum space, and $\ell$ has no relation with an impact parameter any more, but is simply a microscopic length scale present in the two-dimensional theory. The subscript GGRT indicates that with $(D-2)$ bosonic  flavors the $S$-matrix (\ref{GGRT}) describes the worldsheet scattering
of an infinitely long bosonic string propagating in  a $D$-dimensional flat target space and quantized using the light cone quantization \cite{Goddard:1973qh}. Consequently, the classical limit of the theory in this case is described by the Nambu--Goto action,
\be
\label{NG}
S_{NG}=-\ell^2\int d^2\sigma\sqrt{-\det\l{\eta_{\alpha\beta}+\d_\alpha X^i\d_\beta X^i}\r}\;,
\ee
with $i=1,\dots,D-2$. In particular, for $D=26$ the $S$-matrix (\ref{GGRT}) describes worldsheet scattering of a critical bosonic string. Similarly, with an appropriate choice for the number of bosonic and fermionic flavors the same phase shift  (\ref{GGRT}) describes worldsheet scattering of critical superstrings. Let us stress that also at $D\neq 26$ the $S$-matrix (\ref{GGRT}) provides a consistent quantization of (\ref{NG}), which is incompatible, however, with the non-linearly realized target space Poincare symmetry (the $D=3$ case is likely to provide another interesting exception 
\cite{Mezincescu:2010yp}).

 As expected for an $S$-matrix corresponding to a local action, (\ref{GGRT}) exhibits conventional unitarity, analyticity and crossing symmetry properties expected for massless reflectionless scattering \cite{Zamolodchikov:1991vx}. It is also polynomially bounded (in fact, exponentially small) everywhere on the physical sheet, $\mbox{ Im}\, s>0$.\footnote{As usual in a massless theory the amplitude (\ref{GGRT}) exhibits a cut all along the real axis of $s$. Below the cut the amplitude is determined by complex conjugation.}
 
 However, away from the physical sheet the phase shift (\ref{GGRT}) is not polynomially bounded at $s=\infty$ and exhibits an essential singularity there. This singularity appears to be at the origin of many peculiar properties of the shock wave theory (\ref{GGRT}). In particular, it indicates that the $S$-matrix does not correspond to a UV fixed point, and prevents one from extracting local off-shell observables. So the shock wave theory  stands half-way between conventional UV complete quantum field theories and low energy effective field theories requiring UV completion. In the classical limit it is described by a non-renormalizable classical Lagrangian (\ref{NG}), however,  produces finite on-shell amplitudes at the quantum level.  For the purpose of calculating local off-shell observables it does not appear to be better than any non-renormalizable theory.
 We refer to this behavior as asymptotic fragility.
 
 It is worth noting that we are not  claiming that the Lagrangian (\ref{NG}) alone produces finite amplitudes. Presumably, it should be supplemented with an infinite number of scheme dependent counterterms to reproduce (\ref{GGRT}). The lowest order finite counterterm
 in dimensional regularization was calculated in \cite{Dubovsky:2012sh} and is related to the Polchinski--Strominger interaction \cite{Polchinski:1991ax}.
 
 All these properties and the very way how we introduced the shock wave theory here strongly suggest that it should be interpreted as a gravitational theory rather than a conventional quantum field theory. Namely,  it is natural to interpret (\ref{NG}) as a theory of two-dimensional (integrable) gravity coupled to $(D-2)$ free massless bosons. Given that there is no propagating massless spin two particle in two dimensions, one should not be surprised that  coupling to gravity does not change the spectrum, and only modifies (``dresses") scattering amplitudes.
 
To confirm this interpretation one would like to find a prescription to gravitationally dress a broader class of theories.
As was recently pointed out in \cite{Caselle:2013dra}, it is straightforward to generalize the construction to 
an arbitrary CFT.
 A massive generalization of the phase shift (\ref{eikonal}) was pointed out in  \cite{Dubovsky:2012wk}. 

Namely, for two massive particles of a mass $m$  the phase shift
\be
\label{massive}
e^{2i\delta(s)}=e^{i\ell^2\!\sqrt{s(s-4m^2)}/4}
\ee
 exhibits the correct analytic properties and the same asymptotic behavior in the UV as the massless one. 
 We are not aware of a simple closed form for the corresponding classical Lagrangian, but it is straightforward to construct perturbatively  a few of the first terms. 
 For instance, for a single flavor one gets
 \be
 \label{Lagr}
 {\cal L}={1\over 2}(\d\phi)^2-{1\over2} m^2\phi^2+{\ell^2\over 8}\l\l\d\phi\r^4-m^4\phi^4\r+\dots\;.
 \ee
 where $\dots$ stand for terms suppressed by higher powers of $\ell$.
 
\begin{figure}[t]
\begin{center}
\includegraphics[]{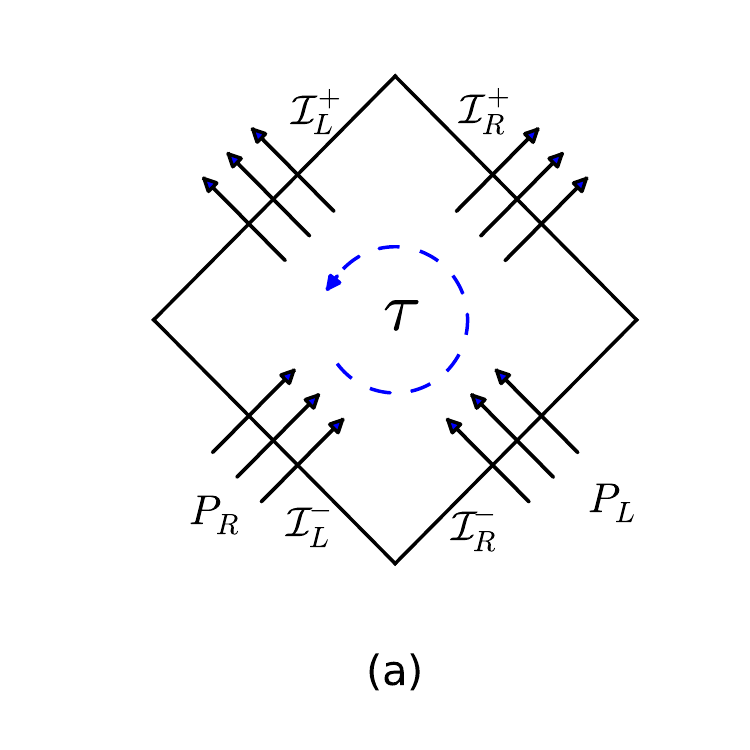}
\includegraphics[]{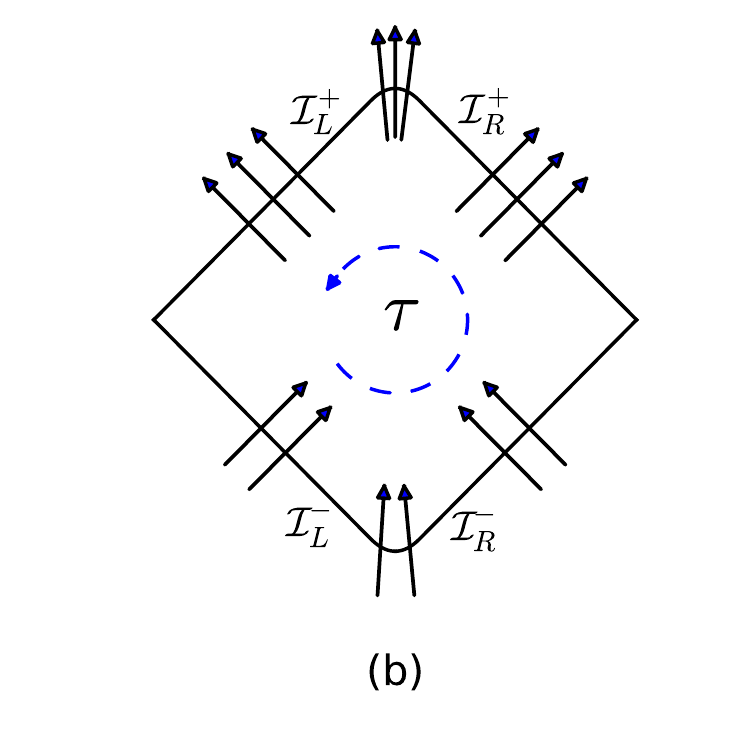}
\caption[]{\small{(a) Null in and out momenta at the holographic boundary of $M_2$ for integrable relativistic shock waves. (b) Generalization to time-like momenta and non-integrable case.} }

\label{boundary}

\end{center}
\end{figure}

 We do not see any direct relation of  the phase shift (\ref{massive}) with higher dimensional gravitational eikonal scattering of massive particles.  However, it turns out that to find its further generalizations it is instructive to follow
 a particular derivation of the relativistic gravitational eikonal amplitude \cite{Verlinde:1991iu}. 
 This derivation results in the following prescription for calculating relativistic eikonal scattering.
 Consider a reparametrization invariant quantum mechanical system of two degrees of freedom $X^\alpha$, $\alpha=0,1$ with the action of  the Chern--Simons form,
 \be
 \label{QMaction}
 S_{CS}[X]=\ell^{-2}\oint d\tau\epsilon_{\alpha\beta}X^\alpha\d_\tau X^\beta\;.
 \ee
 In the original higher dimensional setup these variables were residing on the null boundary. Also there was an additional dependence on the coordinates in the transverse plane.
 For our purposes  the quantum mechanics can be thought of as living at the holographic boundary of the two-dimensional Minkowski space-time, and no transverse coordinates left.
 Then, as a result of the shock wave scattering, the in-state with the set of left- and right-moving null two-momenta $p_{Li}$ and $p_{Rj}$ gets transformed as
\comment{
 \be
 \label{null}
 |\{p_{Ri},p_{Lj}\}\rangle_{in}\to S_0(p_{Ri},p_{Lj})|\{p_{Ri},p_{Lj}\}\rangle_{out}\;,
 \ee}
 \be
 \label{null}
 |\{p_{Ri},p_{Lj}\}\rangle_{out}= S_0(p_{Ri},p_{Lj})|\{p_{Ri},p_{Lj}\}\rangle_{in}\;,
 \ee
 where $S_0$ is given by the quantum mechanical functional integral with simple vertex operator insertions
  \be
 \label{DXnull}
 S_0(p_{Ri},p_{Lj})=\int{\cal D}Xe^{iS_{CS}[X]+i\l\sum_{i}p_{iR}^\alpha X_\alpha(\tau_i)+ \sum_{j}p_{jL}^\alpha X_\alpha(\tau_j)+\sum_{i}\bar{p}_{iR}^\alpha X_\alpha(\bar{\tau}_i)+ \sum_{j}\bar{p}_{jL}^\alpha X_\alpha(\bar{\tau}_j)\r}\;.
 \ee
 Here bars mark out-going particles, and
 \[
 {\bar p}\equiv-p\;.
 \]
 This functional integral is Gaussian and can be evaluated by finding the classical solution for $X^\alpha$ and plugging it back.
 As a consequence of the reparametrization invariance the result depends only on the cyclic order of the insertion times $\tau_i,\tau_j,\bar{\tau}_i,\bar{\tau}_j$. Moreover, nothing changes under arbitrary permutations of insertion times
   within any of these four groups separately. The correct cyclic order to reproduce the amplitude is determined by a physical order of the momenta as they flow through the boundary, see Fig.~\ref{boundary}a,
   \[
   0<\tau_i<\tau_j<\bar{\tau}_i<\bar{\tau}_j<1\;,
   \]
where the boundary time $\tau$ is chosen to be periodic in the interval ${\cal I}=[0,1]$.

It is tempting to identify this simple quantum mechanics with a holographic boundary theory for the integrable gravitational model at hand. However, there are infinitely many ways to represent the exponent in the form of a (functional) integral. To judge whether this identification is useful one would like to see whether it allows for interesting deformations. We will see momentarily that it does.
 
\begin{figure}[t]
\begin{center}
\includegraphics[]{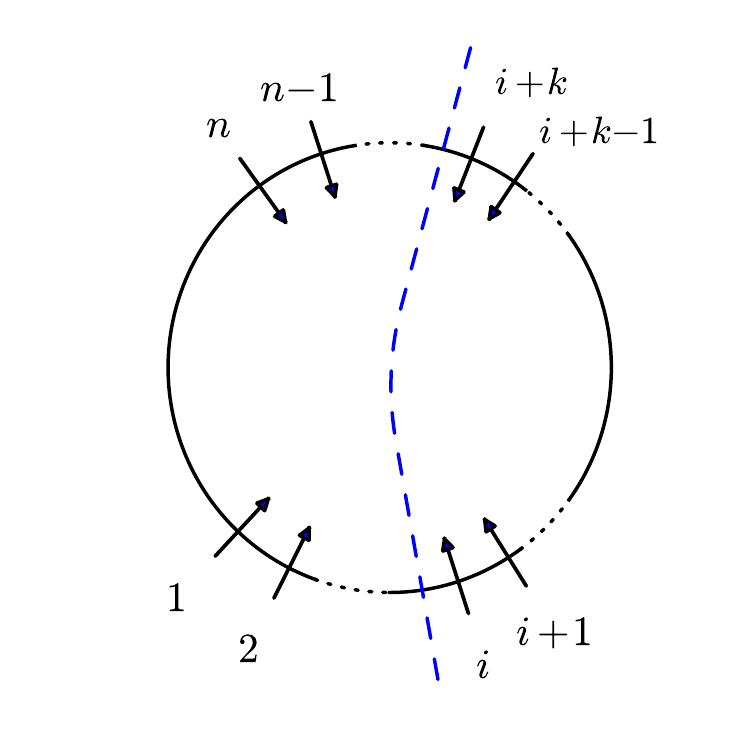}
\caption[]{\small{Minkowskian momenta are naturally cyclically ordered in $2d$. The dressing factor (\ref{Ddress}) factorizes for an arbitrary decomposition of the external momenta into two clusters.} }

\label{fig:RP1}

\end{center}
\end{figure}

\section{Natural Tuning from Gravitational Dressing}
\label{dressing}
As a way to arrive at the recipe for gravitational dressing, let us
 take seriously the boundary quantum mechanics (\ref{QMaction}) as a holographic description of two-dimensional Minkowski gravity, and use it as a guide for extending the set of examples of asymptotically fragile theories. A natural thing to try  is to relax the requirements on the allowed set of particle momenta and to allow for time-like momenta and nonidentical sets of initial and final momenta in the functional integral (\ref{DXnull}).
 To simplify the notations it is convenient to treat all the momenta as incoming, so that particles in a final state carry negative energy. Also we need to extend the cyclic order to cover this more general case. We will do this by applying the same prescription of following the physical order of the momenta through the boundary,  Fig.~\ref{boundary}b, which we used before for the null momenta only. In other words, we simply use the natural cyclic order defined on a set of non-zero two-momenta defined up to a positive rescaling, $p\sim \lambda p$, $\lambda>0$, see Fig.~\ref{fig:RP1}. The order is not well-defined 
for collinear momenta, however, as before, the answer does not change under their permutations. At the end this leads us to consider the following path integral,
\be
\label{DX}
{\cal D}(p_i)=\int{\cal D}Xe^{iS_{CS}[X]+i\sum_{i}p_{i}^\alpha X_\alpha(\tau_i)}\;,
\ee
where momenta are ordered as we just described. As a consequence of the shift symmetry for the Chern-Simons action, $X^\alpha\to X^\alpha+const$, the integral is non-zero only if the total momentum is conserved,
\be
\label{conserve}
\sum_i p_i=0\;.
\ee
As before, the integral is straightforward to evaluate, and the result is
 \be
 \label{Ddress}
 {\cal D}(p_i)=e^{i\ell^2\!/4\sum_{i<j}p_i*p_j}\;,
 \ee
 where ``$*$" stands for the antisymmetric product of the momenta (\ref{skew}). 
Even though the expression (\ref{Ddress}) as written requires an actual order on the set of momenta, it is straightforward to check that it does not change under a cyclic permutation of the momenta, so the cyclic order introduced above is sufficient to make it well-defined.
It is straightforward to check though that this set of functions on its own does not define a consistent set of scattering amplitudes, in particular, unitarity is not satisfied. 

The resolution is to give up a fully holographic description and to start with a conventional
flat space quantum field theory as an input. Then we can use (\ref{Ddress}) not as an amplitude {\it per se}, but rather as a dressing factor modifying the field theory $S$-matrix, according to
\be
\label{dress}
\hat{S}(p_i)={\cal D}(p_i)S(p_i)\;.
\ee
This procedure is reminiscent of the Castillejo--Dalitz--Dyson (CDD) dressing \cite{Castillejo:1955ed} in the context of integrable models. In fact, ${\cal D}(p_i)$ can be thought of as a very special example of the CDD factor.
Its interesting and peculiar property is that it can be used to dress non-integrable $S$-matrices as well.
\subsection{General Properties of Dressed Amplitudes} 
Let us start with  general arguments indicating that gravitational dressing (\ref{dress}) results in a healthy $S$-matrix and then we provide a further support for this with a few illustrative low order perturbative calculations.

Our definition of dressing treats democratically all external momenta, which guarantees that it does not spoil the crossing symmetry of the $S$-matrix. 
It is important to keep in mind that crossing symmetry in two dimensions is less powerful than one is used to in higher dimensional theories. Famously, this allows for the existence of integrable theories, which may exhibit transmissions without reflections and annihilations, and non-trivial three-to-three scattering in the absence of two-to-four processes. Our construction of the amplitude, by making use of the cyclic ordering, also relies on this peculiarity.  

On the other hand, unitarity is not manifest, when the dressed matrix is written in the form (\ref{dress}). Indeed, this is exactly where the construction would fail if we attempted to replace ${\cal D}(p_i)$ with a general CDD factor.
To demonstrate  unitarity of $\hat{S}(p_i)$ let us present it in a slightly different form. As a consequence of the momentum conservation the dressing factor ${\cal D}(p_i)$ has the following property. Let us cut the set of all momenta into two arbitrary adjacent (as defined by cyclic ordering) groups 
$k_a$ and $q_b$ (see Fig.~\ref{fig:RP1}). Then, for all choices of the cut, one finds the following factorized representation for ${\cal D}(p_i)$
\be
\label{cut}
{\cal D}(p_i)=e^{i\ell^2 \!/4 \sum_{a<a'}k_a*k_{a'}}e^{i\ell^2\!/4\sum_{b<b'}q_b*q_{b'}}\;.
\ee 
In particular, if momenta $k_a$ correspond to incoming particles and momenta $q_b$ correspond to outgoing, this implies that the dressed $S$-matrix can be written as a product
\be
\label{USU}
\hat{S}=USU\;,
\ee
where $U$ is a unitary operator defined as
\[
U|\{k_a\}\rangle=e^{i\ell^2\!/4\sum_{a<a'}k_a*k_{a'}}|\{k_a\}\rangle\;.
\]
This proves unitarity of $\hat{S}$.

As far as the analytic structure of $\hat{S}$ goes, it is notoriously hard to precisely formulate all the required analytic properties of multiparticle amplitudes (for introduction see, {\it e.g.},  \cite{Eden}). However, the analytic structure of the dressing factor (\ref{Ddress}) is relatively simple and featureless, so that it appears extremely plausible that dressing does not introduce
any pathologies. Instead of attempting to provide a rigorous proof for this claim, we will try our best to describe the structure of the dressed amplitude as  explicitly as possible, and to illustrate
how this structure arises using simple low order perturbative calculations.

\begin{figure}[t]
\begin{center}
\includegraphics[]{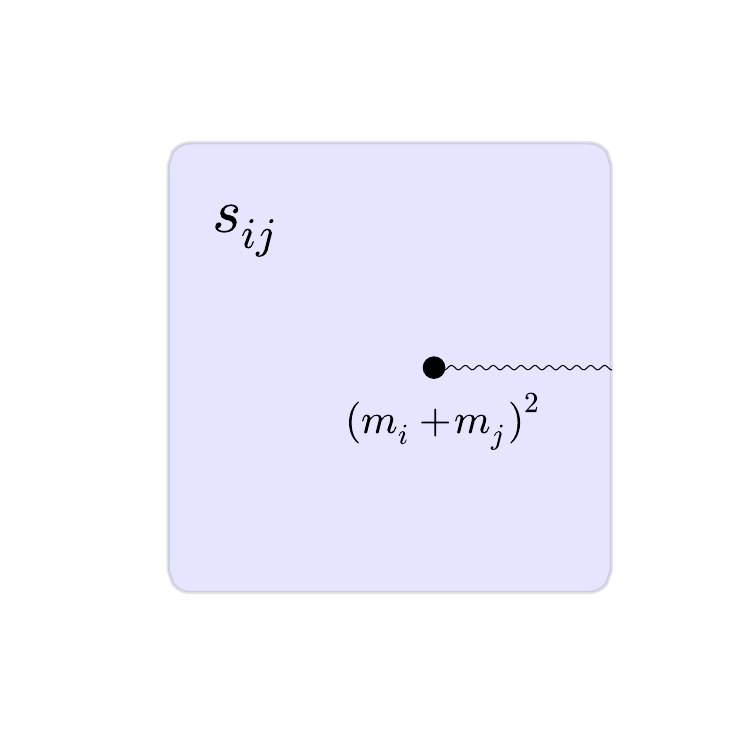}
\caption[]{\small{The physical sheet is defined using the physical two-particle thresholds for all pairs $p_i,p_j$.} }

\label{fig:sij}

\end{center}
\end{figure}

To describe the analytic structure of the dressing factor it is convenient to parametrize the physical sheet in terms of all possible two-particle kinematic invariants
\[
s_{ij}=(p_i+p_j)^2\;.
\]
Note that these are not independent, for example, the two-to-two amplitude in two dimensions is parametrized by a single independent invariant  $s$ and an extra discrete choice, which for equal masses reduces to setting either $t=0$ or $u=0$.
In terms of these invariants one finds,
\be
\label{pij}
p_i*p_j={1\over 2}\sqrt{\l s_{ij}-(m_i+m_j\r^2)\l s_{ij}-(m_i-m_j)^2\r}\;.
\ee
Let us define a physical sheet as the maximal region in the space of complex $s_{ij}$ where one can reach starting from the physical values of the momenta, while remaining on-shell and satisfying  the conservation condition (\ref{conserve}), and without crossing the cuts corresponding to the normal thresholds $s_{ij}=(m_i+m_j)^2$ in any of $s_{ij}$-planes (see Fig.~\ref{fig:sij}). Then the dressing factor is analytic in this whole region. 

This statement is not totally obvious. From (\ref{Ddress}) it is manifest that the dressing factor does not have any poles or essential singularities at finite values of $s_{ij}$. However, given the expression (\ref{pij}), it is not clear that there are no additional cuts on the physical sheet, corresponding  to  anomalous thresholds $s_{ij}=(m_i-m_j)^2$. To see why these are absent, let us parametrize momenta $p_i$ in terms of complex rapidities, which are $2\pi i$ periodic variables defined according to
\[
p_i=m_i(\cosh\theta_i,\sinh\theta_i)\;.
\]
Two-particle kinematic invariants $s_{ij}$ are expressed through the rapidities as
\be
\label{map}
s_{ij}=m_i^2+m_j^2+2 m_im_j\cosh(\theta_i-\theta_j)\;.
\ee
On the physical sheet, that is, away from the cuts shown in Fig. \ref{fig:sij} we have $\Im (\theta_i-\theta_j)\neq 0$ modulo $2\pi i$. Hence, we can cyclically order particle momenta according to the imaginary parts of their rapidities and the dressing factor (\ref{Ddress}) is well defined on the physical sheet.
Note, that this ordering is apparently different from what we used before. So far we were using real parts of rapidities to order particles, and now switched to using imaginary parts. It is straightforward to verify though, that the latter ordering reduces to the former as one approaches the physical regions following the Feynmann $i\epsilon$ prescription.

This argument only proves that the dressing factor is well-defined and analytic on the physical sheet as a function of particle momenta (or, equivalently, rapidities).
However, one can go further and prove that the dressing factor is a well-defined and analytic function on a physical sheet, understood as a complex submanifold in the space of complex $s_{ij}$'s.
Given that $s_{ij}$'s are analytic functions of rapidities, the only non-trivial check to make is that the mapping (\ref{map}) is invertible on the physical sheet, {\it i.e.}  that the values of all $s_{ij}$'s uniquely determine the corresponding particle momenta, up to an overall boost. 

Indeed, as a consequence of cyclic ordering on the physical sheet we have a well-defined notion of neighboring particles. All the momenta are determined
(up to an overall boost) by rapidity differences between neighbors, 
\[
\theta_{i(i+1)}=\theta_{i+1}-\theta_i\;,
\]
$i=1,\dots,n-1$ and the first particle can be chosen arbitrarily.
These rapidity differences are $2\pi i$-periodic variables and restricted to satisfy
\be
\label{thetanb}
0<\Im \theta_{i(i+1)}<\pi\;.
\ee
As a consequence of $2\pi i$-periodicity the first inequality is a matter of definition, while  the second  is a consequence of the
momentum conservation written in the form
\[
\sum m_ie^{i\theta_i}=0\;.
\]
The restriction (\ref{thetanb}) immediately implies that on a physical sheet for any $i$ there is a single value of $\theta_{i(i+1)}$, corresponding to a given $s_{i(i+1)}$, {\it  Q.E.D.}

Finally, as before, the dressing factor is polynomially bounded on the physical sheet, but exhibits the essential singularity at the infinity, which is no worse and no better than what one finds on the string worldsheet.

\subsection{Perturbative Checks}
General axioms of analytic  $S$-matrix theory were always guided by the properties of perturbative amplitudes. So, as a further check for the consistency of the dressed $S$-matrix, let us present several examples illustrating how its  first few terms in the derivative expansion can be reproduced perturbatively from conventional local effective field theory actions.
\subsubsection{Dressing of a Free Massive Field}

Gravitational dressing of a free massive field results in an integrable theory with the  two-to-two $S$-matrix given by (\ref{massive}), which can also be presented  as
\be
\label{2to2}
S=e^{i\ell^2\sqrt{s(s-4m^2)}/4}{\mbf 1}= {\mbf 1}+\frac{i}{2\sqrt{s(s-4m^2)}}\A(s){\mbf 1}.
\ee
Here, we introduced the scattering amplitude $\A(s)$, which is a more conventional object  in the context of perturbation theory.
The unusual normalization factor in (\ref{2to2}) comes from
the Jacobian relating the conventional  energy-momentum conservation delta-function, and the unit operator $\mbf{1}$ acting on two particle states,
\be
\delta^2(p_1+p_2-p_3-p_4)=\frac{2 E_1 E_2}{\sqrt{s(s-4m^2)}}(\delta(\vec p_1-\vec p_3)\delta(\vec p_2-\vec p_4)+\delta(\vec p_1-\vec p_4)\delta(\vec p_2-\vec p_3))\;,
\ee
where $\vec p$ denotes the one-dimensional spatial momentum vector. To make the algebra shorter 
 we often make the discrete choice $t=0$, which corresponds to $\vec p_1=\vec p_3$, $\vec p_2=\vec p_4$. 
 
 The $S$-matrix (\ref{2to2}) is simple enough, so that we can
 explicitly trace which of its properties are most crucial for the possibility of perturbative reconstruction from a local Lagrangian. These are
\begin{enumerate}
\item The existence of a loop counting parameter $\ell$.

\item Crossing symmetry $\A(s+i\eps)=\A(4m^2-s-i\eps)$, which relates the analytic continuation of the $s$-channel amplitude to the $u$-channel amplitude.

\item Unitarity, which through the optical theorem fixes the discontinuities of the $n^{th}$ order amplitude $\A_n(s)=\mathcal{O}(\ell^{2n})$ in terms of the lower order amplitudes.

\end{enumerate}

Let us first expand $\A(s)$ in powers of $\ell^2$. The first term in the expansion,
\[
\A_1=\frac{\ell^2}{2} s(s-4m^2)\;,
\]
 coincides with the tree-level amplitude $\tilde \A_1$ calculated from the Lagrangian \eqref{Lagr}, once the on-shell condition $s+u=4m^2$ is imposed (in the $t=0$ channel). 

Now consider the  order $\ell^4$ amplitude $\A_2=\ell^4 [s(4m^2-s)]^{3/2}/16$ which has branching points at $s=0$ and $s=4m^2$. The discontinuity across the corresponding cuts is completely fixed to be $i\A_1^2/2\sqrt{s(s-4m^2)}$ by unitarity of $\A$. On the other hand, the perturbative 1-loop calculation using vertices $\ell^2(\d\phi)^4$ and $\ell^2m^4\phi^4$ in \eqref{Lagr} must yield an on-shell amplitude $\tilde\A_2$ with identical discontinuity across the $s$- and $u$- channel cuts since $\tilde \A_1=\A_1$. This implies that $\tilde\A_2$ and $\A_2$ differ at most by polynomials in $s$ which can be canceled by adding local quartic counter-terms proportional to $\ell^4$ to $\cal{L}$. 

\begin{figure}[t]
\begin{center}
\includegraphics[scale=0.8]{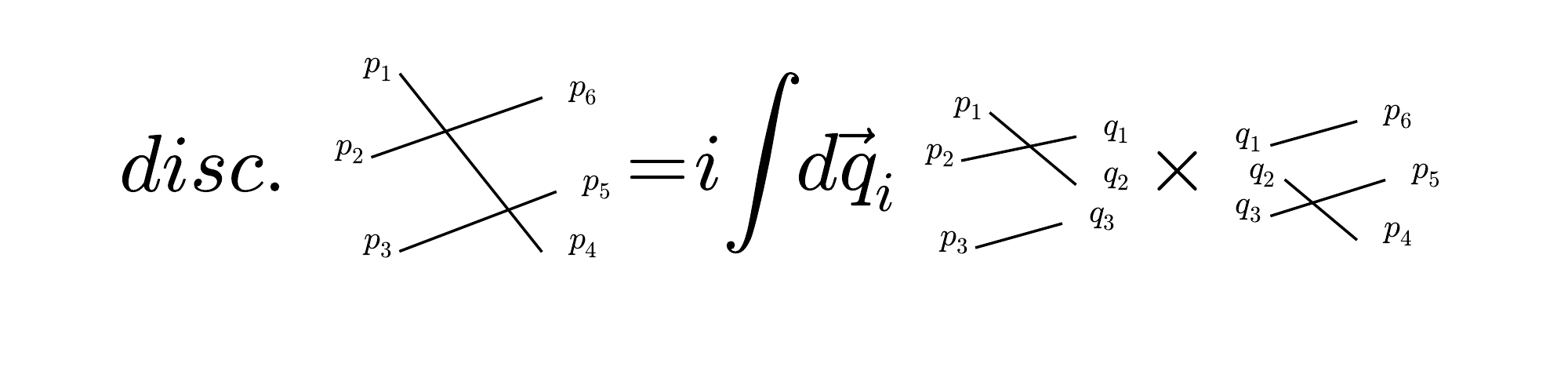}
\caption[]{\small{The discontinuity of the tree-level $3\to 3$ diagram is dictated by the amplitude of the disconnected diagrams and is proportional to $\mbf{1}$ in the momentum space.} }

\label{3to3cut}

\end{center}
\end{figure}

The first connected three-to-three scattering amplitude also appears at order  $\ell^4$. From the factorizable theory \eqref{2to2} one gets a three-to-three $S$-matrix which is proportional to ${\mbf 1}$ in the momentum space. That is, there is one extra delta-function in addition to the energy-momentum conserving $\delta^2(\sum p_i-\sum p_f)$ that naturally appears in the perturbative calculation. Note that this extra delta function is again dictated by unitarity (see Fig.~\ref{3to3cut}) and is generically hidden as the imaginary part of the propagator poles,
\be
\frac{1}{q^2-m^2+i\eps}={\cal{P}}\frac{1}{q^2-m^2}-i\pi\delta(q^2-m^2)\;,
\ee
where $\cal P$ stands for a principal value.
Summing over all diagrams one finds that the principal values cancel up to a polynomial which can again be subtracted by adding sextic counter-terms of order $\ell^4$,\footnote{To reproduce these cancellations it is very helpful to use the parametrization reviewed in \cite{Dorey}.}
\be
\Delta{\cal L}=-\frac{1}{16}\ell^4\left[(\d\phi)^6+\frac{1}{2}m^2\phi^2(\d\phi)^4+\frac{1}{2}m^6\phi^6\right].
\ee
The reason for this cancelations is likely to  be related to  the fact that our two-to-two $S$-matrix (trivially) satisfies the Yang-Baxter equation \cite{Iagolnitzer}.
 
It appears that this construction can be continued inductively to all orders in perturbation theory. Suppose there is a Lagrangian which perturbatively reproduces the scattering amplitude $\A$ up to order $\ell^{2n}$. Unitarity and crossing symmetry fix all discontinuities of the order $\ell^{2(n+1)}$ in the perturbative amplitude $\tilde \A_{n+1}$, and hence the difference $\tilde \A_{n+1}-\A_{n+1}$ would be a polynomial which can be removed by adding $\mathcal{O}(\ell^{2(n+1)})$ counter-terms. 
\subsubsection{1-loop Dressing for $n$ Interacting Fields}
Even though the simple analytic structure of the dressed $S$-matrices 
appears fully consistent and
makes it extremely plausible that in the perturbative regime dressing can always be reconstructed from local Lagrangians, one may worry
that all our examples so far were integrable and that the procedure would fail for non-integrable theories. In fact, integrable theories can be dressed with arbitrary CDD factors, while none of them, except for the gravitational dressing factor (\ref{Ddress}) can be applied to a generic theory where the momentum transfer between particles is allowed.
 In particular, it is very instructive to check explicitly that the restricted 
crossing properties of the dressing factor  are consistent for non-integrable interactions.

Hence, as another  perturbative check, let us see how the dressing works for a non-integrable interaction of 
 $n$ scalar fields with arbitrary masses,
\be
\mathcal{L}_{QFT=} \half\l \sum_i \d \phi_i \d \phi_i - m_i^2 \phi_i^2\r - \lambda \phi_1 \phi_2 \ldots \phi_n\;.
\ee
To simplify the combinatorics we have chosen all the fields in the interaction vertex to be different. 

\begin{figure}[t]
\begin{center}
\includegraphics[]{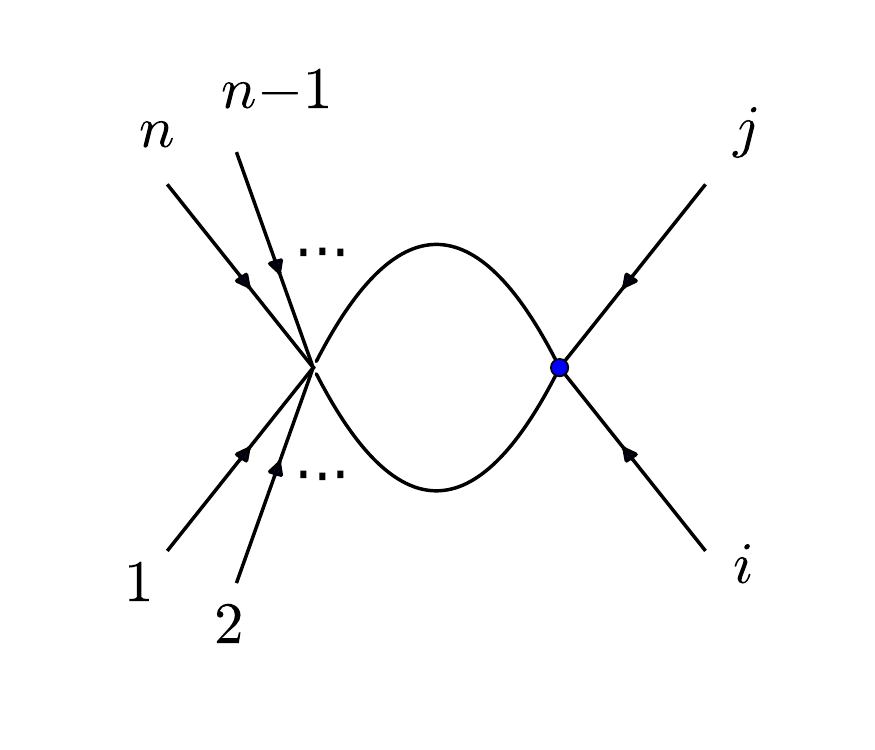}
\caption[]{\small{One-loop diagram describing the dressing of $\lambda\phi^n$ vertex at  order $\ell^2$.} }

\label{fig:phin}

\end{center}
\end{figure}

Already at the leading order in $\lambda$ this theory allows for processes with arbitrary kinematics. The first step is to reproduce the dressing of a free theory with different species at the tree level---that is to the first order in $\ell^2$ and to the zeroth order in $\lambda$. It is straightforward to check that the addition of
\be
\label{Ldr}
\Delta\mathcal{L}_{2}=-\frac{\ell^2}{8} \l( \d_{\alpha}\phi_i  \d^{\alpha}\phi_i)^2-2( \d_{\alpha}\phi_i  \d^{\alpha}\phi_j)^2+m_i^2m_j^2 \phi_i^2\phi_j^2 \r
\ee
does the job at this order by providing the contribution to the two-to-two scattering of particles of type $i$ and $j$ equal to
\be
\frac{i \ell^2}{2} p_i*p_j  \mathbf{1} \; , 
\ee
corresponding to the representation \eqref{cut} of the dressing factor with a cut separating in and out states.
As expected, in the massless limit the coefficients in the action \eqref{Ldr} coincide with those of the Nambu-Goto action.
In the remainder of this section we will demonstrate that the tree-level interactions give rise to the correct one-loop dressing.
Namely, the sum of one-loop graphs represented in Fig.~\ref{fig:phin} 
agrees with the expansion of the dressed scattering $n$-point amplitude,
\be
\label{A1loop}
\A_{1-loop}=\frac{- i\lambda \ell^2}{4} \sum \limits_{i<j}p_i*p_j ,
\ee
up to extra polynomial terms which can be canceled by adding local counterterms to the Lagrangian. Probably the most non-obvious thing to see in this calculation is how the rapidity ordering of Fig.~\ref{fig:RP1} needed to define \eqref{A1loop} arises at the level of Feynman diagrams. Since all the singularities present in the dressing phase are already present at the one-loop order  this calculation provides a strong support for the consistency
of the dressing procedure.

Let us consider the sum of three graphs, corresponding to the three vertices present in \eqref{Ldr}. 
By using the dimensional regularization we get that particles $i$ and $j$ running in the loop contribute to this sum as
\begin{gather}
I_{ij}=-i \lambda \ell^2 \mu^{2-d}\int \frac{d^d q}{(2 \pi)^d} \frac{m_i^2m_j^2+(p_i\cdot p_j)\l(p_i+p_j+q)\cdot q\r}{(q^2-m_i^2)((q+p_i+p_j)^2-m_j^2)} = \frac{\lambda\ell^2\mu^{2-d}}{(4 \pi)^{d/2}}\times \nonumber \\
  \int \limits_0^1 dx\l \Gamma \l 2-\frac{d}{2}\r \Delta^{\frac{d}{2}-2} \l m_i^2m_j^2-(p_i \cdot p_j) (p_i+p_j)^2 x(1-x)\r - \Gamma\l 1-\frac{d}{2}\r \frac{d}{2} \Delta^{\frac{d}{2}-1} p_i \cdot p_j \r\nonumber
\end{gather}
where 
\[
\Delta=-x(1-x)(p_i+p_j)^2+xm_j^2+(1-x)m_i^2\;
\] 
We only need to keep the finite piece of this integral, and can also drop local terms 
which depend on the renormalization scale $\mu$.
It is also convenient to express the result through $\sij=(p_i+p_j)^2$. Then we are left with the following  integral over the Feynman parameter
\bea
\label{Iij}
 I_{ij}\cong \frac{\lambda\ell^2}{8 \pi} \l \int \limits_0^1 dx\frac{(\sij-m_i^2-m_j^2) \sij (x^2-x)+2m_i^2 m_j^2}{\Delta}
+(s_{ij}-m_i^2-m_j^2)  \int \limits_0^1 dx \ln \Delta \r ,
\eea 
where the ``$\cong$" sign is a reminder that we are not keeping track of  local terms.
To account properly for the  $i\epsilon$ prescription let us evaluate the integrals for real $s_{ij}$ in the  region 
\[
(m_i-m_j)^2<s_{ij}<(m_i+m_j)^2\;,
\] 
where both roots of $\Delta$ are complex, so that 
$i\epsilon$ can be set to zero, and then analytically continue to the rest of the physical sheet. It is convenient to integrate the second term in (\ref{Iij}) by parts and combine it with the first one, which gives
\begin{gather}
\label{loopint}
 I_{ij}\cong\frac{\lambda\ell^2}{8 \pi} \int \limits_0^1 dx\frac{2 m_i^2 m_j^2+(\sij-m_i^2-m_j^2)(m_i^2-x \sij)}{\Delta} \cong 
 - \frac{\lambda\ell^2}{8 \pi}R_{ij}
\arctan \l \frac{R_{ij}}{\sij-m_i^2-m_j^2} \r \;
\end{gather}
where
\[
R_{ij}=\sqrt{((m_i+m_j)^2-\sij)(\sij-(m_i-m_j)^2)}\;.
\]
The arctangent is understood to be equal to zero when $s\to(m_i-m_j)^2$ and the square root is positive. This expression gets significantly simplified if one uses the rapidity variables. In terms of these variables $$R_{ij}=-2 i m_i m_j \sinh \thij$$
where $\thij$ is one of the four expressions
 \be
 \label{thchoice}
 \thij=\pm(\theta_i-\theta_j)\; {\mbox{ or }}\; \thij=\pm(\theta_i-\theta_j)+2\pi i\;,
 \ee
 depending on which one of them satisfies $0<-i \thij < \pi$. Then the argument of the arctangent becomes $\tan\l-i\thij \r$, and noticing that $s_{ij}=(m_i-m_j)^2$ at $\thij=i\pi$ the loop integral reduces simply to 
\be
\label{simple}
I_{ij}\cong \frac{\lambda \ell^2 m_i m_j}{4\pi} \l \thij -i\pi \r \sinh \thij\;.
\ee
Now we can analytically continue the result to any point on the physical sheet of $\sij$.
 In the course of this continuation $\thij$  should always stay within the strip $0<\Im\thij\leq\pi$. 
Only one of the four expressions (\ref{thchoice}) belongs to  this strip, unless $\Im \thij=0$ or $\Im \thij=  \pi$. In the first case the choice of the 
sign depends on what side one approaches the cut in the $\sij$ plane (see Fig.~\ref{fig:sij}). In the second case one has two choices, but it is straightforward to see that they are equivalent --- this is the manifestation of analyticity of the function \eqref{loopint} at $\sij=(m_i-m_j)^2$ on the physical sheet. 

 Finally, let us sum \eqref{simple} over all possible pairs $\langle i, j \rangle$. Combining all the terms which multiply any $\theta_i$ one gets an expression vanishing by the energy-momentum conservation,
\be
 \theta_i \frac{\lambda\ell^2}{4\pi} \sum\limits_j m_i m_j \sinh(\theta_i-\theta_j)= 0\;.
\ee 
Consequently, the sum reduces to 
\be
\label{order}
\sum\limits_{\langle i,j \rangle}I_{ij}=\sum\limits_{\langle i,j \rangle} \frac{\lambda\ell^2}{4\pi} \l \thij -i \pi \r m_i m_j \sinh \thij=\sum\limits_{\langle i,j \rangle}\frac{-i \lambda \ell^2}{4} (1-2 \alpha_{ij}) m_i m_j \sinh\l \theta_i-\theta_j \r,
\ee
where the above definition of $\thij$ implies that $\alpha_{ij}=1$ if  $\Im\theta_i< \Im\theta_j$ and zero otherwise. The expression \eqref{order} is ordered with respect to the imaginary parts of rapidities and exactly reproduces \eqref{A1loop}.

\section{Conclusions}
In this concluding section let us first briefly discuss some possible implications of the natural tuning for  particle physics model building, if one takes this scenario seriously.
Note, that this approach may receive a strong observational support if the LHC data reveals additional scalar particles at the electroweak scale 
in the absence of new natural physics protecting their mass. Non-supersymmetric two-Higgs doublet model is a concrete example of such a setup. Indeed, additional
fine-tuning needed to keep extra scalars light is not required by the ``atomic principle",  so that the anthropic explanation will be falsified in this case.

Note also that the restricted formulation of naturalness, as proposed in the Introduction, suggests a restrictive and unconventional set of model building rules.
Within this framework the electroweak scale  becomes a fundamental scale for non-gravitational physics. If there is any model building goal to be solved by introducing new particles directly coupled to the Standard Model  with unsuppressed couplings, these particles cannot have masses much heavier than the electroweak scale without destabilizing the Higgs mass.  On the other hand, unlike in the other two approaches to naturalness,  it is allowed to add new unprotected scalars at the weak scale.

By taking seriously more detailed features of the two-dimensional construction, model building rules may be made extremely restrictive. For instance, for the gravitational dressing to work we had to start with a UV complete quantum field theory. If one literally translates this lesson  into four dimensions, 
then natural tuning implies that  the Standard Model should be embedded into an asymptotically free theory\footnote{Or, possibly, into a theory admitting a strongly coupled UV fixed point.} very close to the electroweak scale in order to get rid of the $U(1)$
Landau pole and preserve field theory naturalness. This
favors exotic weak scale unification scenarios in the spirit of \cite{Dimopoulos:2002mv}. 

We used  the electroweak hierarchy problem as a major reference point in our discussion, but one may wonder whether the cosmological constant problem changes anything. 
On one side, the cosmological constant problem may appear to be in conflict with natural tuning, because already the known Standard Model thresholds contributes to the graviton tadpole diagram, which correspond to the cosmological constant. On the other side,  vacuum energy is not a well-defined field theory observable in the absence of gravity, so that the natural tuning approach, requiring non-gravitational particle physics to be natural, appears to be consistent at least  legalistically.

Furthermore, one might argue that the two-dimensional gravitational dressing actually oversolved the vacuum energy problem---we started with an arbitrary flat space theory and ended up constructing a flat space gravitational theory without any tuning. It would be very interesting to construct AdS/dS generalizations of dressing to understand this issue better. 
Overall, it is fair to say that the anthropic approach remains as the only  comprehensive way to address the cosmological constant problem.

Finally, one may ask whether natural tuning can  fit into string theory. On one side, there is a reason to be optimistic, given that our starting point was the string worldsheet scattering. On the other side, string theory itself 
provides a further example of asymptotic fragility ({\it i.e.}, of non-Wilsonian UV completion, which operates directly at the $S$-matrix level and does not lead to well-defined local off-shell observables), which apparently does not lead to realization of natural tuning. 
In string theory the hierarchy problem arises as a consequence of the landscape at the stage when a choice of the physical vacuum has to be made. 

Moreover, the existence of the landscape  actually promotes the hierarchy problem from a purely aesthetic issue into a sharply-posed physics question within string theory.
Similar to how one can adjust parameters of the materials in the condensed matter lab, string landscape combined with eternal inflation allows for scanning of different quantum field theories in  different places in the Universe. This scanning allows for the anthropic  fine-tuning as a solution to the hierarchy problem, but at the same time also provides a robust physical meaning
for natural solutions. A hierarchy protected by symmetry is  more plausible to be realized because there can be a large island in the landscape, where the corresponding symmetry holds. Note also that anthropic selection  on its own does not provide a preference for tuning against symmetry. It only makes tuning possible, but its actual probability depends on the statistics of vacua and on the dynamics of how they are populated. 

The natural tuning can be realized in this context if there exists a large enough region of the landscape with string coupling of order one, $g_s\sim 1$, 
so that the string and the Planck scales coincide and play the role of the $\ell$ scale above and no new heavy thresholds arise. Whether natural tuning works or not depends on the size of this region in the landscape, the distribution for the values of $\ell$ in this region and on the dynamics for populating vacua with different $\ell$. 

Needless to say, at the moment none of these is known so that  this possibility remains a pure speculation. Still we feel that the construction presented in the paper is non-trivial and present some interest,
at least from the technical point of view and as a concrete illustration of the concept. Applying the principle that ``no stone should be left unturned" when deciding the fate of naturalness, we believe that natural tuning deserves further thought.

\label{conclusions}
\section*{Acknowledgements}
We thank Raphael Flauger for collaboration on many of these topics and for numerous illuminating discussions.
We thank Mina Arvanitaki, Savas Dimopoulos, Giga Gabadadze, Lam Hui, John March-Russell, Joe Polchinski, Sergey Sibiryakov, and Neal Weiner for useful discussions.
This work is supported in part by the NSF grant PHY-1068438. MM was supported by the NASA grant NNX12AF86G S06. SD thanks KITP, Santa Barbara, where this work was completed, for hospitality and partial support through the NSF grant PHY11-25915.
\bibliographystyle{utphys}
\bibliography{dlrrefs}
\end{document}